# Label-free virtual HER2 immunohistochemical staining of breast tissue using deep learning

Bijie Bai [†,1,2,3], Hongda Wang[†,1,2,3], Yuzhu Li[†,1,2,3], Kevin de Haan[1,2,3], Francesco Colonnese[4], Yujie Wan[5], Jingyi Zuo[4], Ngan B. Doan[6], Xiaoran Zhang[1], Yijie Zhang[1,2,3], Jingxi Li[1,2,3], Wenjie Dong[7], Morgan Angus Darrow[8], Elham Kamangar[8], Han Sung Lee[8], Yair Rivenson[1,2,3], and Aydogan Ozcan[*,1,2,3,9]

[1]Electrical and Computer Engineering Department, University of California, Los Angeles, CA, 90095, USA.

[2]Bioengineering Department, University of California, Los Angeles, 90095, USA.

[3]California NanoSystems Institute (CNSI), University of California, Los Angeles, CA, USA.

[4]Computer Science Department, University of California, Los Angeles, CA, USA.

[5]Physics and Astronomy Department, University of California, Los Angeles, CA, 90095, USA

[6]Translational Pathology Core Laboratory, University of California, Los Angeles, CA, 90095, USA

[7] Statistics Department, University of California, Los Angeles, CA, 90095, USA.

[8]Department of Pathology and Laboratory Medicine, University of California at Davis, Sacramento, CA, 95817, USA.

[9]Department of Surgery, University of California, Los Angeles, CA, 90095, USA

[†]Equal contributing authors

[*]Correspondence: Aydogan Ozcan. Email: ozcan@ucla.edu

**Abstract**

The immunohistochemical (IHC) staining of the human epidermal growth factor receptor 2 (HER2) biomarker is widely practiced in breast tissue analysis, preclinical studies and diagnostic decisions, guiding cancer treatment and investigation of pathogenesis. HER2 staining demands laborious tissue treatment and chemical processing performed by a histotechnologist, which typically takes one day to prepare in a laboratory, increasing analysis time and associated costs. Here, we describe a deep learning-based virtual HER2 IHC staining method using a conditional generative adversarial network that is trained to rapidly transform autofluorescence microscopic images of unlabeled/label-free breast tissue sections into bright-field equivalent microscopic images, matching the standard HER2 IHC staining that is chemically performed on the same tissue sections. The efficacy of this virtual HER2 staining framework was demonstrated by quantitative analysis, in which three board-certified breast pathologists blindly graded the HER2 scores of virtually stained and immunohistochemically stained HER2 whole slide images (WSIs) to reveal that the HER2 scores determined by inspecting virtual IHC images are as accurate as their immunohistochemically stained counterparts. A second quantitative blinded study performed by the same diagnosticians further revealed that the virtually stained HER2 images exhibit a comparable staining quality in the level of nuclear detail, membrane clearness, and absence of staining artifacts with respect to their immunohistochemically stained counterparts. This virtual HER2 staining framework bypasses the costly, laborious, and time-consuming IHC staining procedures in laboratory, and can be extended to other types of biomarkers to accelerate the IHC tissue staining used in life sciences and biomedical workflow.

## Introduction

The immunohistochemical (IHC) staining of tissue sections plays a pivotal role in the evaluation process of a broad range of diseases. Since its first implementation in 1941[1], a great variety of IHC biomarkers have been validated and employed in clinical and research laboratories for characterization of specific cellular events[2], e.g., the nuclear protein Ki-67 associated with cell proliferation[3], the cellular tumor antigen P53 associated with tumor formation[4], and the human epidermal growth factor receptor 2 (HER2) associated with aggressive breast tumor development[5]. Due to its capability of selectively identifying targeted biomarkers, IHC staining of tissue has been established as one of the gold standards for tissue analysis and diagnostic decisions, guiding disease treatment and investigation of pathogenesis[6–8].

Though widely used, the IHC staining of tissue still requires a dedicated laboratory infrastructure and skilled operators (histotechnologists) to perform laborious tissue preparation steps and is therefore time-consuming and costly. Recent years have seen rapid advances in deep learning-based virtual staining techniques, providing promising alternatives to the traditional histochemical staining workflow by computationally staining the microscopic images captured from label-free thin tissue sections, bypassing the laborious and costly chemical staining process. Such label-free virtual staining techniques have been demonstrated using autofluorescence imaging[9,10], quantitative phase imaging[11], light scattering imaging[12], among others[13–15], and have successfully created multiple types of histochemical stains, e.g., hematoxylin and eosin (H&E)[9–14], Masson's trichrome[9–11], and Jones silver stains[9–11]. These previous works did not perform any virtual IHC staining and mainly focused on the generation of structural tissue staining, which enhances the contrast of specific morphological features in tissue sections. In a related line of research, deep learning has also enabled the prediction of biomarker status (e.g., Ki-67 quantification[16]) and tumor prognostic from H&E-stained microphotographs of various malignancies including hepatocellular carcinoma[17], breast cancer[18–22], bladder cancer[23], thyroid cancer[24,25], and melanoma[26]. These studies highlight a possible correlation between the presence of specific biomarkers and morphological microscopic changes in the tissue; however, they do not provide an alternative to IHC stained tissue images that reveal sub-cellular biomarker information for pathologists' diagnostic inspection for e.g., inter- and intra-cellular signatures such as cytoplasmic and nuclear details[27].

Here, we present a deep learning-based label-free virtual IHC staining method (Fig. 1), which transforms autofluorescence microscopic images of unlabeled tissue sections into bright-field equivalent images, matching the standard IHC stained images of the same tissue samples. In this study, we specifically focused on the IHC staining of HER2, which is an important cell surface receptor protein that is involved in regulating cell growth and differentiation[28,29]. Assessing the level of HER2 expression in breast tissue, i.e., HER2 status, is routinely practiced based on the HER2 IHC staining of the formalin-fixed, paraffin-embedded (FFPE) tissue sections, and helps predict the prognosis of breast cancer and its response to HER2-directed immunotherapies[5,29–33]. For example, the intracellular and extracellular studies of HER2 have led to the development of pharmacological anti-HER2 agents that benefit the treatment of HER2-positive tumors[34–38]. Further efforts are being made to develop new pharmacological solutions that can counter HER2-directed-drug resistance and improve treatment outcomes in clinical trials[39–42]. With numerous animal models established for preclinical studies and life sciences related research, a deeper understanding of the oncogene, biological functionality, and drug resistance mechanisms of HER2 is being explored[43–47]. In addition to these, HER2 biomarker was also used as an essential tool in developing and testing of novel biomedical imaging[48,49], statistics[50], and spatial transcriptomics[51] methods.

The presented virtual HER2 staining method is based on a deep learning-enabled image-to-image transformation, using a conditional generative adversarial network (GAN), as shown in Fig. 2. Once the training phase was completed, two blinded quantitative studies were performed using new breast tissue sections with different HER2 scores to demonstrate the efficacy of our virtual HER2 staining framework. For this purpose, we used the semi-quantitative Dako HercepTest scoring system[52], which involves assessing the percentage of tumor cells that exhibit membranous staining for HER2 along with the intensity of the staining. The results are reported as 0 (negative), 1+ (negative), 2+ (weakly positive/equivocal), and 3+ (positive). In the first study, three board-certified breast pathologists blindly graded the HER2 scores of virtually stained HER2 whole slide images (WSIs) as well as their IHC stained standard counterparts. Our results and the statistical analysis revealed that determining the HER2 status based on our virtual HER2 WSIs is as accurate as standard analysis based on the chemically-prepared IHC HER2 slides. In the second study, the same pathologists rated the staining quality of both virtual HER2 and standard IHC HER2 images using different metrics, i.e., nuclear detail, membrane clearness, background staining, and staining artifacts. This study revealed that at least two pathologists out of the three agreed that there is no statistically significant difference between the virtual HER2 staining image quality and the standard IHC HER2 staining image quality in the level of nuclear detail, membrane clearness, and absence of staining artifacts.

The presented framework achieved the first demonstration of label-free virtual IHC staining, and bypasses the costly, laborious, and time-consuming IHC staining procedures that involve toxic chemical compounds. This virtual HER2 staining technique has the potential to be extended to virtual staining of other biomarkers and may accelerate the IHC-based tissue analysis workflow in life sciences and biomedical applications, while also enhancing the repeatability and standardization of IHC staining.

## Results

**Label-free virtual HER2 staining of breast tissue**

We demonstrated our virtual HER2 staining method by training deep neural network (DNN) models with a dataset of 25 breast tissue sections collected from 19 unique patients, constituting in total 20,910 image patches, each with 1024×1024 pixels. Once a DNN model was trained, it virtually stained the unlabeled tissue sections using their autofluorescence microscopic images captured with DAPI, FITC, TxRed, and Cy5 filter cubes (see Methods section), matching the corresponding bright-field images of the same field-of-views, captured after standard IHC HER2 staining. In the network training and evaluation process, we employed a cross-validation approach. Separate network models were trained with different dataset divisions to generate 12 virtual HER2 WSIs for blind testing, i.e., 3 WSIs at each of the 4 HER2 scores (0, 1+, 2+, and 3+). Each virtual HER2 WSI corresponds to a unique patient that was not used during the network training phase. Note that all the tissue sections were obtained from existing tissue blocks, where the HER2 reference (ground truth) scores were provided by UCLA Translational Pathology Core Laboratory (TPCL) under UCLA IRB 18-001029.

Fig. 3 summarizes the comparison of the virtual HER2 images inferred by our DNN models against their corresponding IHC HER2 images captured from the same tissue sections after standard IHC staining. Both the WSIs and the zoomed-in regions show a high degree of agreement between virtual staining and standard IHC staining. These results indicate that a well-trained virtual staining network can reliably transform the autofluorescence images of unlabeled breast tissue sections into the bright-field equivalent, virtual HER2 images, which match their IHC HER2 stained counterparts, across all the HER2 statuses, 0, 1+, 2+, and 3+. Upon close examination, our board-certified pathologists confirmed that the comparison between the IHC and virtual HER2 images showed equivalent staining with no significant perceptible differences in intracellular features such as membrane clarity or nuclear details. In particular, the virtual staining network clearly produced the expected intensity and distribution of membranous HER2 staining (or lack thereof) in tumor cells. In HER2 positive (3+, Figs. 3a-e) breast cancers, both virtually stained and IHC stained images showed strong complete membranous staining in >10% of tumor cells, as well as dim cytoplasmic staining in tumor cells. None of the stromal and inflammatory cells showed false positive staining and the nuclear details of the tumor cells were comparable in both panels. In equivocal (2+, Figs. 3f-j) tumors, virtual images showed weak to moderate membranous staining in >10% of tumor cells, providing the same amount of membranous staining of tumor cells in corresponding areas. HER2 negative (1+, Figs. 3k-o) tumors showed faint membranous staining in 10% or more of tumor cells. None of the stromal and inflammatory cells showed faint staining. HER2 negative (0, Figs. 3p-t) tumor showed no staining in the tumor cells.

**Blind evaluation and quantification of virtual HER2 staining**

Next, we evaluated the efficacy of the presented virtual HER2 staining framework with a quantitative blinded study in which the 12 virtual HER2 WSIs and their corresponding standard IHC HER2 WSIs were mixed and presented to three board-certified breast pathologists who graded the HER2 score (i.e., 3+, 2+, 1+, or 0) for each WSI without knowing if the image was from a virtual stain or standard IHC stain. Random image shuffling, rotation, and flipping were applied to the WSIs to promote blindness in evaluations. The HER2 scores of the virtual and the standard IHC WSIs that were blindly graded by the three pathologists are summarized in Fig. 4 and compared to their reference, ground truth scores provided by UCLA TPCL. The confusion matrices of virtual HER2 WSIs (Fig. 4a) and IHC HER2 WSIs (Fig. 4b), each corresponding to $N=36$ evaluations, reveal that our virtual HER2 staining approach achieved a similar level of accuracy for HER2 status assessment as the standard IHC staining. Close examination of these confusion matrices reveals that the sum of the diagonal elements of the virtual HER2-based evaluations (22) is higher than that of the IHC HER2 (19), showing that more cases were correctly scored based on virtual HER2 WSIs compared to those based on standard IHC HER2 WSIs. Furthermore, the sum of the absolute off-diagonal errors of virtual HER2-based evaluations (14) is smaller than that of the standard IHC HER2 (18). Based on the same confusion matrices shown in Fig. 4, a chi-square test was performed to compare the degree of agreement between virtual staining and standard IHC staining methods in HER2 scoring. The test results indicate that there is no statistically significant difference between the two methods ($P=0.4752$, see Supplementary Table 1).

In addition to evaluating the efficacy of virtual staining in HER2 scoring, we also quantitatively evaluated the staining quality of the virtual HER2 images and compared them to the standard IHC HER2 images. In this blinded study, we randomly extracted 10 regions-of-interest (ROIs) from each of the 12 virtual HER2 WSIs and 10 ROIs at the same locations from each of their corresponding IHC HER2 WSIs, building a test set of 240 image patches. Each image patch has $8000\times8000$ pixels ($1.3\times1.3$ mm$^2$), which was also randomly shuffled, rotated, and flipped before being reviewed by the same three pathologists. These pathologists were asked to grade the image quality of each ROI based on four pre-designated feature metrics for HER2 staining: membrane clearness, nuclear detail, absence of excessive background staining, and absence of staining artifacts (Fig. 5). The grade scale for each metric is from 1 to 4, with 4 representing perfect, 3 representing very good, 2 representing acceptable, and 1 representing unacceptable. Fig. 5a summarizes the staining quality scores of virtual HER2 and standard IHC HER2 images based on our pre-defined feature metrics, which were averaged over all image patches and pathologists. Figs. 5b-e further compare the average quality scores at each of the 4 HER2 statuses under each feature metric. In Fig. 5b, the membrane clearness scores of HER2 negative ROIs are noted as "not applicable" since there is no staining of the cell membrane in HER2 negative samples. It is important to emphasize that, the standard IHC HER2 images had an advantage in these comparisons because they were pre-selected: a significant percentage of the standard IHC HER2 tissue slides suffered from unacceptable staining quality issues (see Discussion and Supplementary Fig. 1), and therefore they were excluded from our comparative studies in the first place. Nevertheless, the quality scores of virtual and standard IHC HER2 staining are very close to each other and fall within their standard deviations (dashed lines in Fig. 5). We also performed one-sided t-tests on each feature metric evaluated by board-certified pathologists to determine whether standard IHC HER2 images are statistically significantly better than the virtual HER2 images in staining quality. The t-test results showed that only for the metric of 'absence of excessive background staining', two of the three pathologists reported a statistically significant improvement in the quality of the standard IHC staining compared to the virtual staining. For the rest of the feature metrics (i.e., nuclear details, membrane clearness, and staining artifacts), at least two of the three pathologists reported that the staining quality of the IHC HER2 images is not statistically significantly better than their virtual HER2 counterparts (Supplementary Table 2). Also note that the virtually stained HER2 images did not mislead the diagnosis at the whole slide level as also analyzed using the confusion matrices shown in Fig. 4 and the chi-square test reported in Supplementary Table 1.

Besides rating the staining quality of each ROI, the pathologists also graded a HER2 score for each ROI, the results of which are reported in Supplementary Fig. 2. Each histogram in Supplementary Fig. 2a summarizes the HER2 scores of the 10 ROIs extracted from each WSI evaluated by 3 pathologists (i.e., $N$=30 evaluations). The reference (ground truth) HER2 scores of the corresponding WSIs are plotted as gray dashed lines. This analysis reveals that, for the majority of the patients, there is no discrepancy between HER2 scores evaluated from virtually generated ROIs and standard IHC stained ROIs. For the cases where there is a disagreement (e.g., Patients #5 and #11), the histograms of the virtual HER2 scores were centered closer to the reference HER2 scores (dashed lines) compared to the histograms of the standard IHC-based HER2 scores. It is important to also note that grading the HER2 scores from sub-sampled ROIs vs. from the WSI can yield different results due to the inhomogeneous nature of the tissue sections.

**Discussion**

We demonstrated a deep learning-enabled label-free virtual IHC staining method. By training a DNN model, our method generated virtual HER2 images from the autofluorescence images of unlabeled tissue sections, matching the bright-field images captured after standard IHC-staining. Compared to chemically performing the IHC staining, our virtual HER2 staining method is rapid and simple to operate. The conventional IHC HER2 staining involves laborious sample treatment steps demanding a histotechnologist's periodic monitoring (see Supplementary Note 1), and this whole process typically takes one day before the slides can be reviewed by diagnosticians. In contrast, the presented virtual HER2 staining method bypasses these laborious and costly steps, and generates the bright-field equivalent HER2 images computationally using the autofluorescence images captured from label-free tissue sections. After the training is complete (which is a one-time effort), the entire inference process using a virtual staining network only takes ~12 seconds for 1 mm$^2$ of tissue using a consumer-grade computer, which can be further improved by using faster hardware acceleration units.

Another advantage of the presented method is its capability of generating highly consistent and repeatable staining results, minimizing the staining variations that are commonly observed in standard IHC staining. The IHC HER2 staining procedure is delicate and laborious as it requires accurate control of time, temperature, and concentrations of the reagents at each tissue treatment step; in fact, it often fails to generate satisfactory stains. In our study, ~30% of the sample slides were discarded because of unsuccessful standard IHC staining and/or severe tissue damage even though the IHC staining was performed by accredited pathology labs. Supplementary Fig. 1 shows two examples of the standard IHC staining failures we experienced, including complete tissue damage and false negative staining that failed to reflect the correct HER2 score. In contrast, our computational virtual staining approach does not rely on the chemical processing of the tissue and generates reproducible results, which is important for the standardization of the HER2 interpretation by eliminating commonly experienced staining variations and artifacts.

Since the autofluorescence input images of tissue slices were captured with standard filter sets installed on a conventional fluorescence microscope, the presented approach is ready to be implemented on existing fluorescence microscopes without hardware modifications or customized optical components. Our results showed that the combination of the four commonly used fluorescence filters (DAPI, FITC, TxRed, and Cy5) provided a very good baseline for the virtual HER2 staining performance. As an ablation study, we also quantitatively compared virtual staining networks that are trained with different autofluorescence input channels by calculating peak signal-to-noise ratio (PSNR) and structural similarity index (SSIM)[53] between the network output and ground truth images (see Supplementary Fig. 3). Since the staining of the cell membrane is an important assessment factor in HER2 status evaluation, we also performed color deconvolution[54] to split out the membrane stain channel (i.e., diaminobenzidine, DAB stain) followed by calculating and comparing the SSIM scores (Supplementary Fig. 4). These analyses revealed that the performance of the virtual staining network partially degraded with decreasing number of input autofluorescence channels, motivating the use of DAPI, FITC, TxRed, and Cy5 altogether (Supplementary Fig. 3b).

The success of our virtual HER2 staining method relies on the processing of the complex spatial-spectral information that is encoded in the autofluorescence images of label-free tissue using convolutional neural networks. The presented virtual staining method can potentially be expanded to a wide range of other IHC stains. Though our virtual HER2 staining framework was demonstrated based on autofluorescence imaging of unlabeled tissue sections, other label-free microscopy modalities may also be utilized for this task, such as holography[11], fluorescence lifetime imaging[55,56] and Raman microscopy[57]. In addition to generalizing to other types of IHC stains in the assessment of various biomarkers, this method can be further adapted to non-fixed fresh tissue samples or frozen sections, which can potentially provide real-time virtual IHC images for intraoperative consultation during surgical operations.

To the best of our knowledge, this is the first demonstration of label-free virtual IHC staining, and we believe that it opens up new avenues for various applications in life sciences and biomedical diagnostics and can potentially transform the traditional IHC staining workflow.

## Methods

### Sample preparation and standard IHC staining

The unlabeled breast tissue blocks were provided by the UCLA TPCL under UCLA IRB 18-001029 and were cut into 4 μm thin sections. The FFPE thin sections were then deparaffinized and covered with glass coverslips. After acquiring the autofluorescence microscopic images, the unlabeled tissue sections were sent to accredited pathology labs for standard IHC HER2 staining, which was performed by UCLA TPCL and the Department of Anatomic Pathology of Cedars-Sinai Medical Center in Los Angeles, USA. The IHC HER2 staining protocol provided by UCLA TPCL is described in Supplementary Note 1.

### Image data acquisition

The autofluorescence images of the unlabeled tissue sections were captured using a standard fluorescence microscope (IX-83, Olympus) with a ×40/0.95NA (UPLSAPO, Olympus) objective lens. Four fluorescent filter cubes, including DAPI (Semrock DAPI-5060C-OFX, EX 377/50 nm, EM 447/60 nm), FITC (Semrock FITC-2024B-OFX, EX 485/20 nm, EM 522/24 nm), TxRed (Semrock TXRED-4040C-OFX, EX 562/40 nm, EM 624/40 nm), and Cy5 (Semrock CY5-4040C-OFX, EX 628/40 nm, EM 692/40 nm) were used to capture the autofluorescence images at different excitation-emission wavelengths. Each autofluorescence image was captured with a scientific complementary metal-oxide-semiconductor (sCMOS) image sensor (ORCA-flash4.0 V2, Hamamatsu Photonics) with an exposure time of 150 ms, 500 ms, 500 ms, and 1000 ms for DAPI, FITC, TxRed, and Cy5 filters, respectively. The image acquisition process was controlled by μManager (version 1.4) microscope automation software[58]. After the standard IHC HER2 staining is complete, the bright-field WSIs were acquired using a slide scanner microscope (AxioScan Z1, Zeiss) with a ×20/0.8NA objective lens (Plan-Apo).

**Image preprocessing and registration**

The matching of the autofluorescence (network input) and the bright-field IHC HER2 (network ground truth) image pairs is critical for the successful training of an image-to-image transformation network. The image processing workflow for preparing the training dataset for our virtual HER2 staining network is described in Supplementary Fig. 5, which was implemented in MATLAB (MathWorks). First, the autofluorescence images (before the IHC staining) and the whole-slide bright-field images (after the IHC staining) of the same tissue sections were stitched into WSIs (Supplementary Fig. 5a) and globally co-registered by detecting and matching the speeded up robust features (SURF) points[59] (Supplementary Fig. 5b). Then, these coarsely matched autofluorescence and bright-field WSIs were cropped into pairs of image tiles of 1024×1024 pixels (Supplementary Fig. 5c). These image pairs were not accurately matched at the pixel level due to optical aberrations and morphological changes of the tissue structure during the standard (laborious) IHC staining procedures. In order to calculate the transformation between the autofluorescence image and its bright-field counterpart using a correlation-based elastic registration algorithm[60], a registration model[9] needs to be trained to match the style of the autofluorescence images to the style of the bright-field images (Supplementary Fig. 5d). This registration network used the same architecture as our virtual staining network. Following the image style transformation using the registration network (Supplementary Fig. 5e), the pyramid elastic image registration algorithm[60,61] was performed to hierarchically match the local features of the sub-image blocks and calculate the transformation maps. The transformation maps were then applied to correct for the local wrappings of the ground truth images (Supplementary Fig. 5f) which were then better matched to their autofluorescence counterparts. This training-registration process (Supplementary Fig. 5d-f) was repeated 3-5 times until the autofluorescence input and the bright-field ground truth image patches were accurately matched at the single pixel-level (Supplementary Fig. 5g). At last, a manual data cleaning process was performed to remove image pairs with artifacts such as tissue-tearing (during the standard chemical staining process) or defocusing (during the imaging process).

**Virtual HER2 staining network architecture and training schedule**

In this work, a GAN-based network model[62] was employed to perform the transformation from the 4-channel label-free autofluorescence images (DAPI, FITC, TxRed, and Cy5) to the corresponding bright-field virtual HER2 images, as shown in Fig. 2. This GAN framework includes (1) a generator network that creates virtually stained HER2 images by learning the statistical transformation between the input autofluorescence images and the corresponding bright-field IHC stained HER2 images (ground truth), and (2) a discriminator network that learns to discriminate the virtual HER2 images created by the generator from the actual IHC stained HER2 images. The generator and the discriminator were alternatively optimized and simultaneously improved through this competitive training process. Specifically, the generator ($G$) and discriminator ($D$) networks were optimized to minimize the following loss functions:

$$l_{\text{generator}} = \alpha \times L_1\{I_{\text{target}}, G(I_{\text{input}})\} - \lambda \times \log((1 + SSIM\{I_{\text{target}}, G(I_{\text{input}})\})/2) + \gamma \times BCE\left\{D\left(G(I_{\text{input}})\right), 1\right\}$$

$$l_{\text{discriminator}} = BCE\left\{D\left(G(I_{\text{input}})\right), 0\right\} + BCE\{D(I_{\text{target}}), 1\}$$

where $G(\cdot)$ represents the generator inference, $D(\cdot)$ represents the probability of being a real, actually-stained IHC image predicted by the discriminator, $I_{\text{input}}$ denotes the input label-free autofluorescence images, and $I_{\text{target}}$ denotes the ground truth, standard IHC stained image. The coefficients ($\alpha, \lambda, \gamma$) in $l_{\text{generator}}$ were empirically set as (10, 0.2, 0.5) to balance the pixel-wise smooth $L_1$ error[63] of the generator output with respect to its ground truth, SSIM loss[53] of the generator output, and the binary cross-entropy (BCE) loss of the discriminator predictions of the output image. Compared to using the mean squared error (MSE) loss, the smooth $L_1$ loss is a robust estimator that prevents exploding gradients by using MSE around zero and mean absolute error (MAE) in other parts[64]. Specifically, smooth $L_1$ loss between two images $A$ and $B$ is defined as:

$$L_1\{A, B\} = \frac{1}{M \times N}\left(\sum_{\substack{m,n \\ |A(m,n)-B(m,n)|<\beta}} 0.5 \times \frac{(A(m,n) - B(m,n))^2}{\beta} + \sum_{\substack{m,n \\ |A(m,n)-B(m,n)|\geq\beta}} |A(m,n) - B(m,n)| - 0.5\beta\right)$$

where $m$ and $n$ are the pixel indices, the $M \times N$ represents the total number of pixels in each image. $\beta$ was set to 1 in our case.

The SSIM of two images is defined as[53]:

$$SSIM\{A, B\} = \frac{(2\mu_A\mu_B + c_1)(2\sigma_{AB} + c_2)}{(\mu_A^2 + \mu_B^2 + c_1)(\sigma_A^2 + \sigma_B^2 + c_2)}$$

where $\mu_A$ and $\mu_B$ are the mean values of the images $A$ and $B$, $\sigma_A^2$ and $\sigma_B^2$ are the variance of images $A$ and $B$, and $\sigma_{AB}$ is the covariance between images $A$ and $B$. $c_1$ and $c_2$ were set to be $0.01^2$ and $0.03^2$, respectively[53].

The BCE with logits loss used in our network is defined as:

$$BCE\{p, q\} = -[q \times \log(\text{sigmoid}(p)) + (1 - q) \times \log(1 - \text{sigmoid}(p))]$$

where $p$ represents the discriminator predictions and $q$ represents the actual labels (0 or 1).

As shown in Fig. 2a, the generator network was built following the attention U-Net architecture[65] with 4 resolution levels, which can map the label-free autofluorescence images into the HER2 stained images by learning the transformations of spatial features at different spatial scales, catching both the high-resolution local features at shallower levels and the larger scale global context at deeper levels. Our attention U-Net structure is composed of a down-sampling path and an up-sampling path that are symmetric to each other. The down-sampling path contains four down-sampling convolutional blocks, each consisting of a two-convolutional-layer residual block, followed by a leaky rectified linear unit[66] (Leaky ReLU) with a slope of 0.1, and a 2×2 max-pooling operation with a stride size of 2 for down-sampling. The two-convolutional-layer residual blocks contain two consecutive convolutional layers with a kernel size of 3×3 and a convolutional residual path[67] connecting the in and out tensors of the two convolutional layers. The numbers of the input channels and the output channels at each level of the down-sampling path were set to 4, 64, 128, 256, and 64, 128, 256, 512, respectively.

Symmetrically, the up-sampling path contains four up-sampling convolutional blocks with the same design as the down-sampling convolutional blocks, except that the 2× down-sampling operation was replaced by a 2× bilinear up-sampling operation. The input of each up-sampling block is the concatenation of the output tensor from the previous block with the corresponding feature maps at the matched level of the down-sampling path passing through the attention gated connection. An attention gate consists of three convolutional layers and a sigmoid operation, which outputs an activation weight map highlighting the salient spatial features[65]. The numbers of the input channels and the output channels at each level of the up-sampling path were 1024, 1024, 512, 256, and 1024, 512, 256, 128, respectively. Following the up-sampling path, a two-convolutional layer residual block together with another single convolutional layer reduces the number of channels to 3, matching that of our ground truth images (i.e., 3-channel RGB images). Additionally, a two-convolutional-layer center block was utilized to connect and match the dimensions of the down-sampling path and the up-sampling path.

The structure of the discriminator network is illustrated in Fig. 2b. An initial block containing one convolutional layer followed by a Leaky ReLU operation first transformed the 3-channel generator output or ground truth image to a 64-channel tensor. Then, five successive two-convolutional-layer residual blocks were added to perform 2× down-sampling and expand the channel numbers of each input tensor. The 2× down-sampling was enabled by setting the stride size of the second convolutional layer in each block as 2. After passing through the five blocks, the output tensor was averaged and flattened to a one-dimensional vector, which was then fed into two fully connected layers to obtain the probability of the input image being the standard IHC-stained image.

The full image dataset contains 25 WSIs from 19 unique patients, making a set of 20,910 image patches, each with a size of 1024×1024 pixels. For the training of each virtual staining model used in our cross-validation studies, the dataset was divided as follows: (1) Test set: images from the WSIs of 1-2 unique patients (~10%, not overlapped with training or validation patients); after splitting out the test set, the remaining WSIs were further divided to (2) Validation set: images from 2 of the WSIs (~10%), and (3) Training set: images from the remaining WSIs (~80%). The network models were optimized using image patches of 256×256 pixels, which were randomly cropped from the images of 1024×1024 pixels in the training dataset. An Adam optimizer with weight decay[68] was used to update the learnable parameters at a learning rate of $1\times10^{-4}$ for the generator network and $1\times10^{-5}$ for the discriminator network, with a batch size of 28. The generator/discriminator update frequency was set to 2:1. Finally, the best model was selected based on the best MSE loss, assisted with the visual assessment of the validation images. The networks converged after ~120 hours of training.

**Implementation details**

The image preprocessing was implemented in MATLAB using version R2018b (MathWorks). The virtual staining network was implemented using Python version 3.9.0 and Pytorch version 1.9.0. The training was performed on a desktop computer with an Intel Xeon W-2265 central processing unit (CPU), 64 GB random-access memory (RAM), and an Nvidia GeForce GTX 3090 graphics processing unit (GPU).

**Blind evaluation of HER2 images**

For the evaluation of WSIs, 24 high-resolution WSIs were randomly shuffled, rotated, and flipped, and uploaded to an online image viewing platform that was shared with three board-certified pathologists to blindly evaluate and score the HER2 status of each WSI using the Dako HercepTest scoring system[52]. For the evaluation of sub-ROI images, the 240 image patches were randomly shuffled, rotated, and flipped, and uploaded to an online image sharing platform GIGAmacro (https://www.gigamacro.com/). The pathologists' blinded assessments are provided in Supplementary Data 1.

**Statistical analysis**

A chi-square test (two-sided) was performed to compare the agreement of the HER2 scores evaluated based on the virtual staining and the standard IHC staining. Paired t-tests (one-sided) were used to compare the image quality of virtual staining vs. standard IHC staining. We first calculated the differences between the scores of the virtual and IHC image patches cropped from the same positions, i.e., subtracted the score of each IHC stained image from the score of the corresponding virtually stained image. Then one-sided t-tests were performed to compare the differences with 0, by each feature metric and each pathologist (see the Supplementary Information). For all tests, a $P$ value of $\leq 0.05$ was considered statistically significant. All the analyses were performed using SAS v9.4 (The SAS Institute, Cary, NC).

# References

1. Coons, A. H., Creech, H. J. & Jones, R. N. Immunological Properties of an Antibody Containing a Fluorescent Group. *Proceedings of the Society for Experimental Biology and Medicine* **47**, 200–202 (1941).

2. Whiteside, G. & Munglani, R. TUNEL, Hoechst and immunohistochemistry triple-labelling: an improved method for detection of apoptosis in tissue sections—an update. *Brain Research Protocols* **3**, 52–53 (1998).

3. Scholzen, T. & Gerdes, J. The Ki-67 protein: From the known and the unknown. *Journal of Cellular Physiology* **182**, 311–322 (2000).

4. Surget, S., Khoury, M. P. & Bourdon, J.-C. Uncovering the role of p53 splice variants in human malignancy: a clinical perspective. *Onco Targets Ther* **7**, 57–68 (2013).

5. Mitri, Z., Constantine, T. & O'Regan, R. The HER2 Receptor in Breast Cancer: Pathophysiology, Clinical Use, and New Advances in Therapy. *Chemother Res Pract* **2012**, 743193 (2012).

6. Ramos-Vara, J. A. & Miller, M. A. When Tissue Antigens and Antibodies Get Along: Revisiting the Technical Aspects of Immunohistochemistry—The Red, Brown, and Blue Technique. *Vet Pathol* **51**, 42–87 (2014).

7. Ramos-Vara, J. A. Technical Aspects of Immunohistochemistry. *Vet Pathol* **42**, 405–426 (2005).

8. Rojo, M. G., Bueno, G. & Slodkowska, J. Review of imaging solutions for integrated quantitative immunohistochemistry in the Pathology daily practice. *Folia Histochemica et Cytobiologica* **47**, 349–354 (2009).

9. Rivenson, Y. *et al.* Deep learning-based virtual histology staining using auto-fluorescence of label-free tissue. *Nat Biomed Eng* **3**, 466–477 (2019).

10. Zhang, Y. *et al.* Digital synthesis of histological stains using micro-structured and multiplexed virtual staining of label-free tissue. *Light Sci Appl* **9**, 78 (2020).

11. Rivenson, Y. *et al.* PhaseStain: the digital staining of label-free quantitative phase microscopy images using deep learning. *Light Sci Appl* **8**, 23 (2019).

12. Ryu, S. *et al.* Label-free histological imaging of tissues using Brillouin light scattering contrast. *Biomed. Opt. Express, BOE* **12**, 1437–1448 (2021).

13. Chen, Z., Yu, W., Wong, I. H. M. & Wong, T. T. W. Deep-learning-assisted microscopy with ultraviolet surface excitation for rapid slide-free histological imaging. *Biomed. Opt. Express, BOE* **12**, 5920–5938 (2021).

14. Pradhan, P. *et al.* Computational tissue staining of non-linear multimodal imaging using supervised and unsupervised deep learning. *Biomed. Opt. Express, BOE* **12**, 2280–2298 (2021).

15. Li, J. *et al.* Biopsy-free in vivo virtual histology of skin using deep learning. *Light Sci Appl* **10**, 233 (2021).

16. Liu, Y. *et al.* Predict Ki-67 Positive Cells in H&E-Stained Images Using Deep Learning Independently From IHC-Stained Images. *Frontiers in Molecular Biosciences* **7**, (2020).

17. Chen, M. *et al.* Classification and mutation prediction based on histopathology H&E images in liver cancer using deep learning. *npj Precis. Onc.* **4**, 1–7 (2020).

18. Naik, N. *et al.* Deep learning-enabled breast cancer hormonal receptor status determination from base-level H&E stains. *Nat Commun* **11**, 5727 (2020).

19. Couture, H. D. *et al.* Image analysis with deep learning to predict breast cancer grade, ER status, histologic subtype, and intrinsic subtype. *npj Breast Cancer* **4**, 1–8 (2018).

20. Bychkov, D. *et al.* Deep learning identifies morphological features in breast cancer predictive of cancer ERBB2 status and trastuzumab treatment efficacy. *Sci Rep* **11**, 4037 (2021).

21. Binder, A. *et al.* Morphological and molecular breast cancer profiling through explainable machine learning. *Nat Mach Intell* **3**, 355–366 (2021).

22. Shamai, G. *et al.* Artificial Intelligence Algorithms to Assess Hormonal Status From Tissue Microarrays in Patients With Breast Cancer. *JAMA Network Open* **2**, e197700 (2019).

23. Xu, H. *et al. Spatial heterogeneity and organization of tumor mutation burden and immune infiltrates within tumors based on whole slide images correlated with patient survival in bladder cancer*. 554527 https://www.biorxiv.org/content/10.1101/554527v5 (2020) doi:10.1101/554527.

24. Dolezal, J. M. *et al.* Deep learning prediction of BRAF-RAS gene expression signature identifies noninvasive follicular thyroid neoplasms with papillary-like nuclear features. *Mod Pathol* **34**, 862–874 (2021).

25. Anand, D. *et al.* Weakly supervised learning on unannotated H&E-stained slides predicts BRAF mutation in thyroid cancer with high accuracy. *J Pathol* **255**, 232–242 (2021).

26. Kim, R. H. *et al. A Deep Learning Approach for Rapid Mutational Screening in Melanoma*. 610311 https://www.biorxiv.org/content/10.1101/610311v2 (2020) doi:10.1101/610311.

27. Connolly, J. L. *et al.* Role of the Surgical Pathologist in the Diagnosis and Management of the Cancer Patient. *Holland-Frei Cancer Medicine. 6th edition* (2003).

28. Rubin, I. & Yarden, Y. The basic biology of HER2. *Annals of Oncology* **12**, S3–S8 (2001).

29. Iqbal, N. & Iqbal, N. Human Epidermal Growth Factor Receptor 2 (HER2) in Cancers: Overexpression and Therapeutic Implications. *Molecular Biology International* **2014**, 1–9 (2014).

30. Ross, J. S. & Fletcher, J. A. The HER-2/neu Oncogene in Breast Cancer: Prognostic Factor, Predictive Factor, and Target for Therapy. *The Oncologist* **3**, 237–252 (1998).

31. Bilous, M. *et al.* Current Perspectives on HER2 Testing: A Review of National Testing Guidelines. *Mod Pathol* **16**, 173–182 (2003).

32. Burstein, H. J. The Distinctive Nature of HER2-Positive Breast Cancers. *http://dx.doi.org/10.1056/NEJMp058197* https://www.nejm.org/doi/10.1056/NEJMp058197 (2009) doi:10.1056/NEJMp058197.

33. Borg, Å. *et al.* HER-2/neu Amplification Predicts Poor Survival in Node-positive Breast Cancer. *Cancer Res* **50**, 4332–4337 (1990).

34. Costa, R. L. B. & Czerniecki, B. J. Clinical development of immunotherapies for HER2+ breast cancer: a review of HER2-directed monoclonal antibodies and beyond. *npj Breast Cancer* **6**, 1–11 (2020).

35. Wang, J. & Xu, B. Targeted therapeutic options and future perspectives for HER2-positive breast cancer. *Sig Transduct Target Ther* **4**, 1–22 (2019).

36. Pinto, A. C., Ades, F., Azambuja, E. de & Piccart-Gebhart, M. Trastuzumab for patients with HER2 positive breast cancer: Delivery, duration and combination therapies. *The Breast* **22**, S152–S155 (2013).

37. Hudelist, G. *et al.* Her-2/neu-triggered intracellular tyrosine kinase activation: in vivo relevance of ligand-independent activation mechanisms and impact upon the efficacy of trastuzumab-based treatment. *Br J Cancer* **89**, 983–991 (2003).

38. Agus, D. B. *et al.* Phase I Clinical Study of Pertuzumab, a Novel HER Dimerization Inhibitor, in Patients With Advanced Cancer. *JCO* **23**, 2534–2543 (2005).

39. Kaushik Tiwari, M. *et al.* Direct targeting of amplified gene loci for proapoptotic anticancer therapy. *Nat Biotechnol* 1–10 (2021) doi:10.1038/s41587-021-01057-5.

40. Ni, J. *et al.* Combination inhibition of PI3K and mTORC1 yields durable remissions in mice bearing orthotopic patient-derived xenografts of HER2-positive breast cancer brain metastases. *Nat Med* **22**, 723–726 (2016).

41. Kang, J. C. *et al.* Engineering a HER2-specific antibody–drug conjugate to increase lysosomal delivery and therapeutic efficacy. *Nat Biotechnol* **37**, 523–526 (2019).

42. Singh, J. C., Jhaveri, K. & Esteva, F. J. HER2-positive advanced breast cancer: optimizing patient outcomes and opportunities for drug development. *Br J Cancer* **111**, 1888–1898 (2014).

43. Creedon, H. *et al.* Use of a genetically engineered mouse model as a preclinical tool for HER2 breast cancer. *Disease Models & Mechanisms* **9**, 131–140 (2016).

44. Fry, E. A., Taneja, P. & Inoue, K. Clinical applications of mouse models for breast cancer engaging HER2/neu. *Integr Cancer Sci Ther* **3**, 593–603 (2016).

45. De Giovanni, C. *et al.* Vaccines against human HER2 prevent mammary carcinoma in mice transgenic for human HER2. *Breast Cancer Res* **16**, R10 (2014).

46. Piechocki, M. P., Ho, Y.-S., Pilon, S. & Wei, W.-Z. Human ErbB-2 (Her-2) transgenic mice: a model system for testing Her-2 based vaccines. *J Immunol* **171**, 5787–5794 (2003).

47. Jordan, N. V. *et al.* HER2 expression identifies dynamic functional states within circulating breast cancer cells. *Nature* **537**, 102–106 (2016).

48. Giesen, C. *et al.* Highly multiplexed imaging of tumor tissues with subcellular resolution by mass cytometry. *Nat Methods* **11**, 417–422 (2014).

49. Glenn, D. R. *et al.* Single cell magnetic imaging using a quantum diamond microscope. *Nat Methods* **12**, 736–738 (2015).

50. Hafner, M., Niepel, M., Chung, M. & Sorger, P. K. Growth rate inhibition metrics correct for confounders in measuring sensitivity to cancer drugs. *Nat Methods* **13**, 521–527 (2016).

51. Vickovic, S. *et al.* High-definition spatial transcriptomics for in situ tissue profiling. *Nat Methods* **16**, 987–990 (2019).

52. Herceptest[TM] Interpretation Manual - Breast Cancer. https://www.agilent.com/cs/library/usermanuals/public/28630_herceptest_interpretation_manual-breast_ihc_row.pdf.

53. Wang, Z., Bovik, A. C., Sheikh, H. R. & Simoncelli, E. P. Image Quality Assessment: From Error Visibility to Structural Similarity. *IEEE Trans. on Image Process.* **13**, 600–612 (2004).

54. Landini, G., Martinelli, G. & Piccinini, F. Colour deconvolution: stain unmixing in histological imaging. *Bioinformatics* **37**, 1485–1487 (2021).

55. Lakowicz, J. R., Szmacinski, H., Nowaczyk, K., Berndt, K. W. & Johnson, M. Fluorescence Lifetime Imaging. *Anal Biochem* **202**, 316–330 (1992).

56. Bastiaens, P. I. H. & Squire, A. Fluorescence lifetime imaging microscopy: spatial resolution of biochemical processes in the cell. *Trends in Cell Biology* **9**, 48–52 (1999).

57. Andersen, M. E. & Muggli, R. Z. Microscopical techniques in the use of the molecular optics laser examiner Raman microprobe. *Anal. Chem.* **53**, 1772–1777 (1981).

58. Computer Control of Microscopes Using μManager - Edelstein - 2010 - Current Protocols in Molecular Biology - Wiley Online Library. https://currentprotocols.onlinelibrary.wiley.com/doi/full/10.1002/0471142727.mb1420s92.

59. Bay, H., Ess, A., Tuytelaars, T. & Van Gool, L. Speeded-Up Robust Features (SURF). *Computer Vision and Image Understanding* **110**, 346–359 (2008).

60. Saalfeld, S., Fetter, R., Cardona, A. & Tomancak, P. Elastic volume reconstruction from series of ultra-thin microscopy sections. *Nat Methods* **9**, 717–720 (2012).

61. Wang, H. *et al.* Deep learning enables cross-modality super-resolution in fluorescence microscopy. *Nat Methods* **16**, 103–110 (2019).

62. Isola, P., Zhu, J.-Y., Zhou, T. & Efros, A. A. Image-to-Image Translation with Conditional Adversarial Networks. in *2017 IEEE Conference on Computer Vision and Pattern Recognition (CVPR)* 5967–5976 (IEEE, 2017). doi:10.1109/CVPR.2017.632.

63. Huber, P. J. Robust Estimation of a Location Parameter. *The Annals of Mathematical Statistics* **35**, 73–101 (1964).

64. Willmott, C. J. & Matsuura, K. Advantages of the mean absolute error (MAE) over the root mean square error (RMSE) in assessing average model performance. *Climate Research* **30**, 79–82 (2005).

65. Oktay, O. *et al.* Attention U-Net: Learning Where to Look for the Pancreas. 10.

66. Maas, A. L., Hannun, A. Y. & Ng, A. Y. Rectifier Nonlinearities Improve Neural Network Acoustic Models. 6.

67. He, K., Zhang, X., Ren, S. & Sun, J. Deep Residual Learning for Image Recognition. in *2016 IEEE Conference on Computer Vision and Pattern Recognition (CVPR)* 770–778 (IEEE, 2016). doi:10.1109/CVPR.2016.90.

68. Kingma, D. P. & Ba, J. Adam: A Method for Stochastic Optimization. *arXiv:1412.6980 [cs]* (2017).

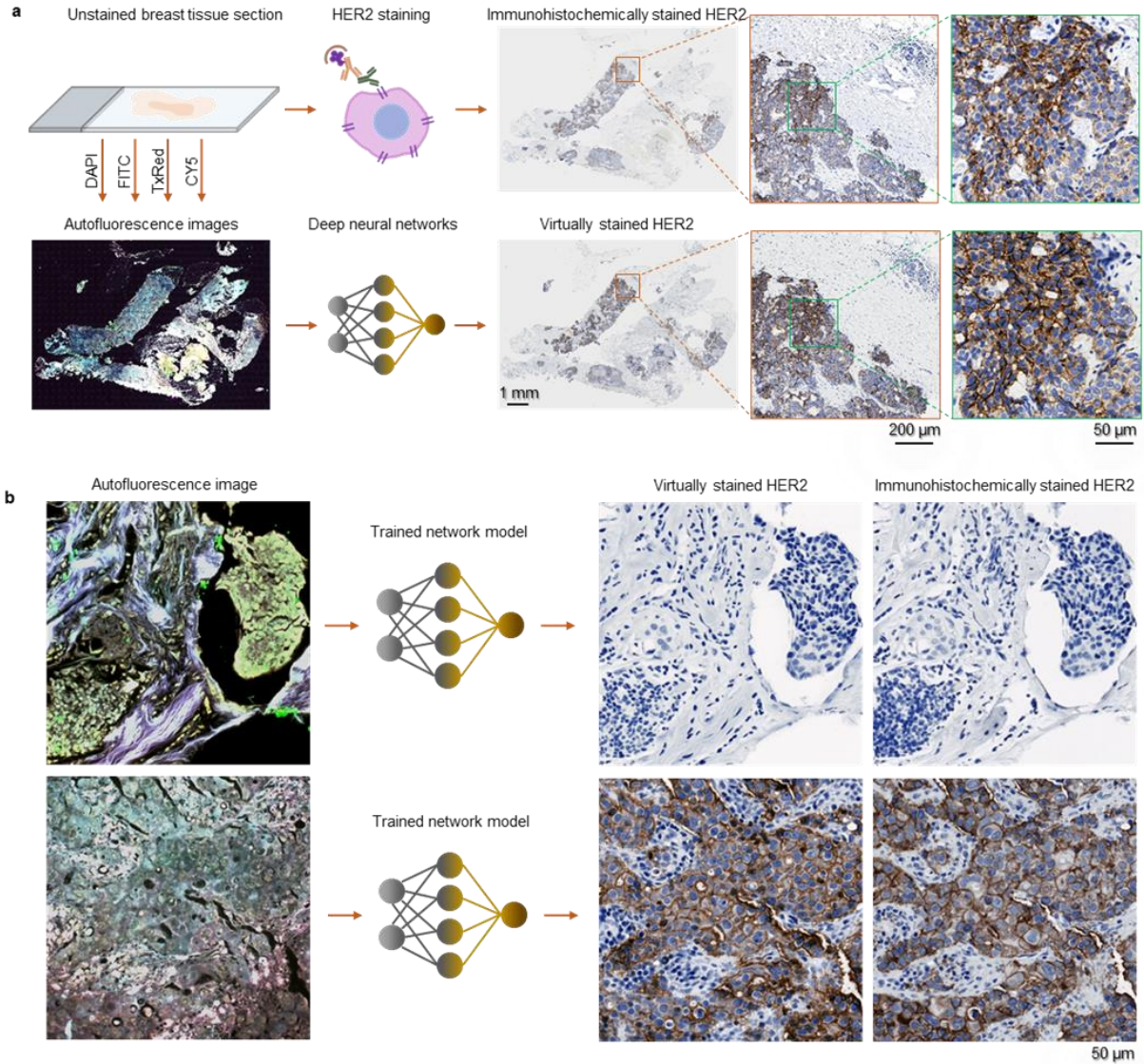

**Figure 1. Virtual HER2 staining of unlabeled tissue sections via deep learning. a,** The standard immunohistochemical (IHC) HER2 staining (top) relies on tedious and costly tissue processing performed by histotechnologists, which typically takes ~1 day. A pre-trained deep neural network enables virtual HER2 staining of unlabeled tissue sections (bottom). **b,** Virtual HER2 staining transforms autofluorescence images of unlabeled tissues sections into bright-field equivalent images that match the images of standard IHC HER2 staining.

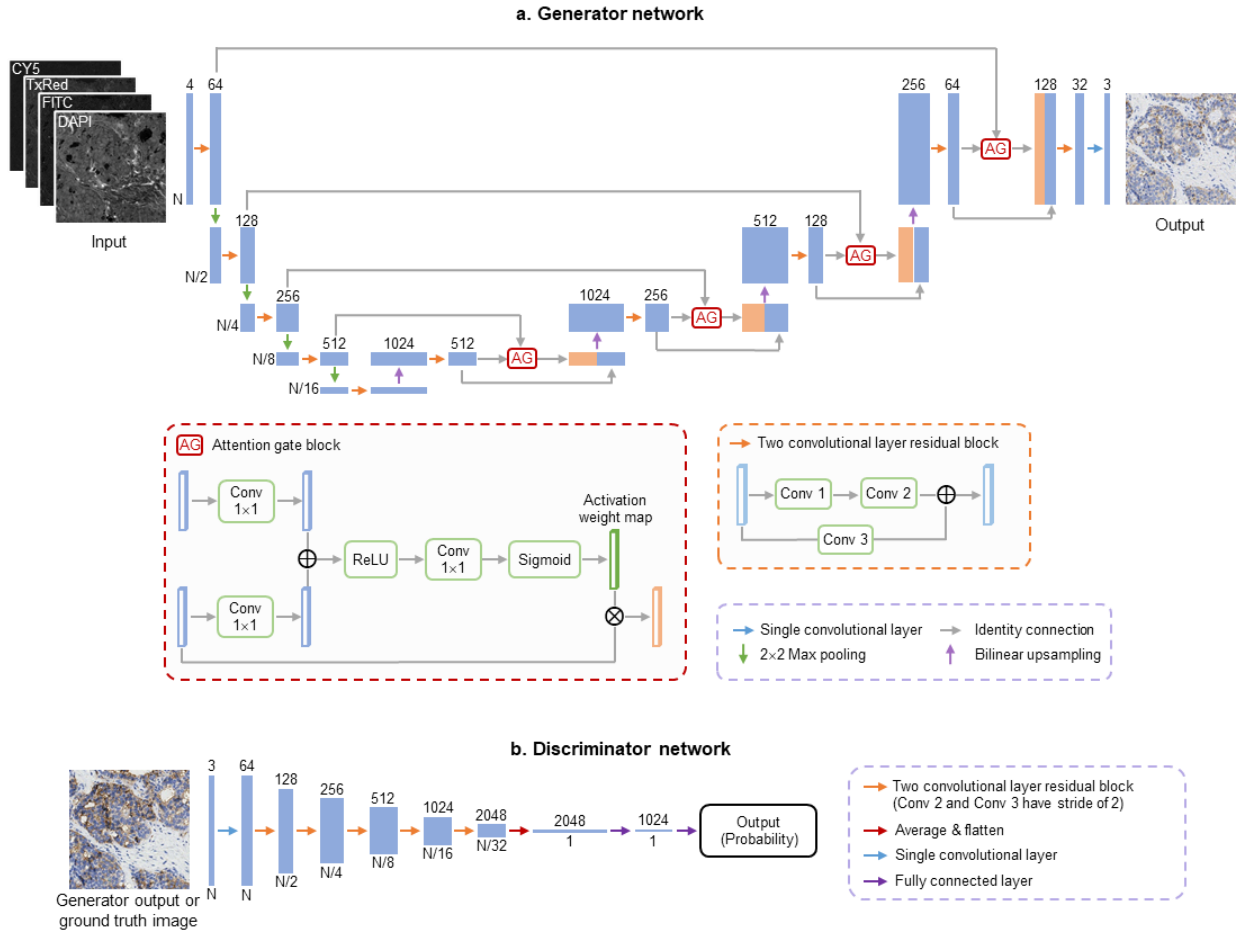

**Figure 2. Virtual HER2 staining network.** A GAN framework which consists of a generator model and a discriminator model was used to train the virtual HER2 staining network. **a,** The generator uses an attention-gated U-net structure to map the label-free autofluorescence images into bright-field equivalent HER2 images. **b,** The discriminator is a CNN composed of five successive two-convolutional-layer residual blocks and two fully connected layers (see Methods). Once the network models converge, only the generator model is used to infer the virtual HER2 images, which takes ~12 seconds for 1 mm$^2$ of tissue area.

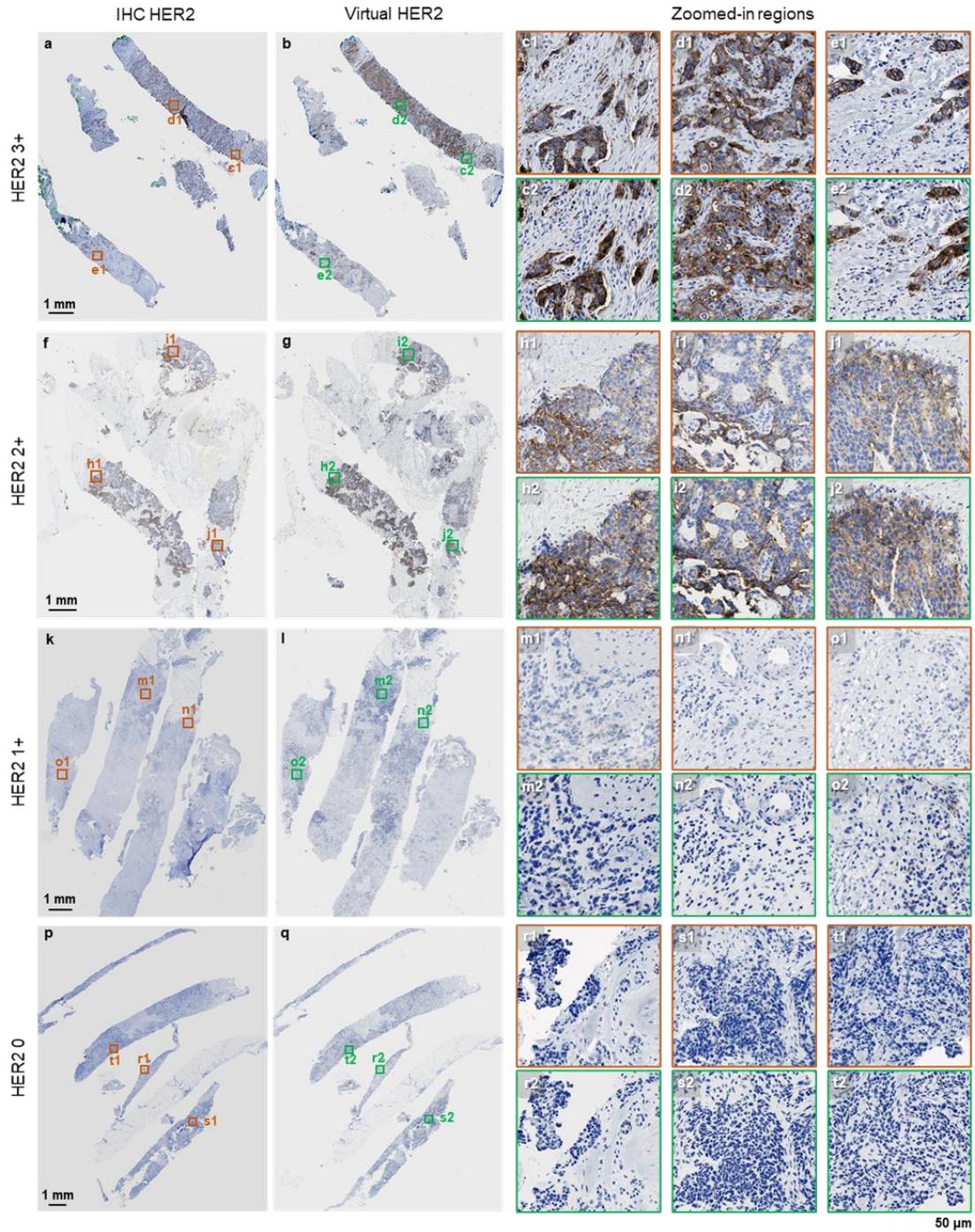

**Figure 3. Comparison of virtual and standard IHC HER2 staining of breast tissue sections at different HER2 scores. a, f, k, p,** Bright-field WSIs of standard IHC HER2 stained samples at **a** HER2 3+, **f** HER2 2+, **k** HER2 1+, and **p** HER2 0. **b, g, l, q,** Bright-field WSIs generated by virtual staining, corresponding to the same samples as **a, f, l, p** respectively. **c1-e1, c2-e2,** Zoomed-in regions of interest from **a, b** at a HER2 score of 3+. **h1-j1, h2-j2,** Zoomed-in regions of interest from **f, g** at a HER2 score of 2+. **m1-o1, m2-o2,** Zoomed-in regions of interest from **k, l** at a HER2 score of 1+. **r1-t1, h2-t2,** Zoomed-in regions of interest from **p, q** at a HER2 score of 0.

## a. Virtual HER2 staining confusion matrix

| | | Reference HER2 score | | | |
|---|---|---|---|---|---|
| | | 3+ | 2+ | 1+ | 0 |
| Evaluated HER2 score (Virtual staining) | 3+ | 9 | 5 | 0 | 0 |
| | 2+ | 0 | 4 | 3 | 0 |
| | 1+ | 0 | 0 | 6 | 6 |
| | 0 | 0 | 0 | 0 | 3 |

Sum of diagonal elements = 22 (max = 36)
Sum of off-diagonal errors = 14

## b. IHC HER2 staining confusion matrix

| | | Reference HER2 score | | | |
|---|---|---|---|---|---|
| | | 3+ | 2+ | 1+ | 0 |
| Evaluated HER2 score (Chemical staining) | 3+ | 3 | 2 | 0 | 0 |
| | 2+ | 6 | 7 | 0 | 1 |
| | 1+ | 0 | 0 | 3 | 2 |
| | 0 | 0 | 0 | 6 | 6 |

Sum of diagonal elements = 19 (max = 36)
Sum of off-diagonal errors = 18

**Figure 4. Confusion matrices of HER2 scores.** Each element in the matrices represents the number of WSIs with their HER2 scores evaluated by board-certified pathologists (rows) based on: **a** virtual HER staining or **b** standard IHC HER2 staining, compared to the reference (ground truth) HER2 scores provided by UCLA TPCL (columns).

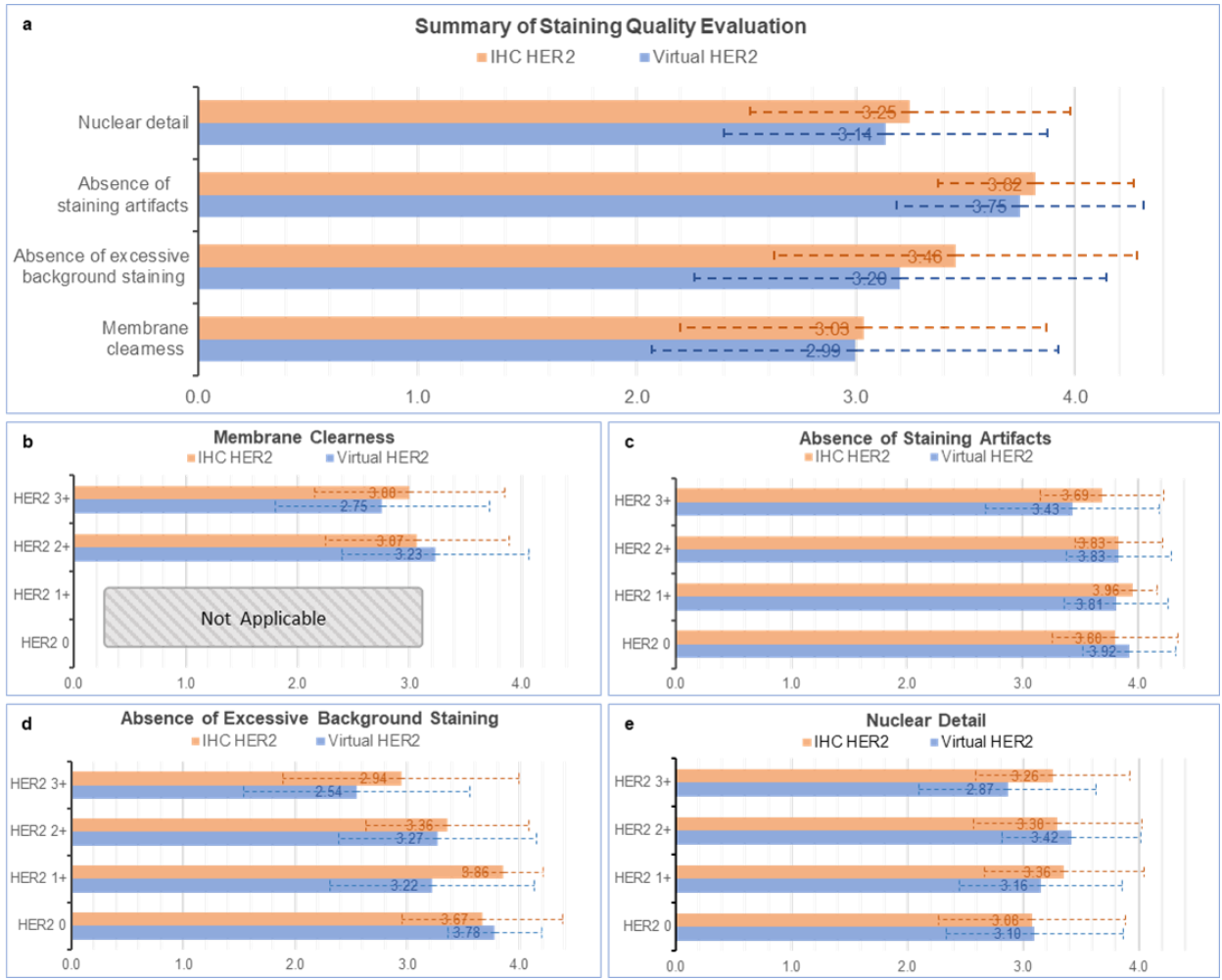

**Figure 5. Comparisons of image quality of virtual HER2 and standard IHC HER2 staining. a,** Quality scores of virtual HER2 and standard IHC HER2 images calculated based on 4 different feature metrics: nuclear details, absence of staining artifacts, absence of excessive background staining, and membrane clearness. Each value was averaged over all the image patches and pathologists. **b-e,** Detailed comparisons of quality scores under each feature metric at different HER2 scores. The grade scale applied for each metric is 1 to 4: 4 for perfect, 3 for very good, 2 for acceptable, and 1 for unacceptable. The standard deviations are plotted by dashed lines.